\documentclass[fleqn,usenatbib]{mnras}

\usepackage{newtxtext,newtxmath,float}

\usepackage[T1]{fontenc}

\DeclareRobustCommand{\VAN}[3]{#2}
\let\VANthebibliography\thebibliography
\def\thebibliography{\DeclareRobustCommand{\VAN}[3]{##3}\VANthebibliography}

\usepackage{graphicx}	
\usepackage{amsmath}	
\usepackage{natbib}
\usepackage{url}
\usepackage{longtable}
\usepackage{threeparttable}
\usepackage{multirow}
\usepackage{booktabs}
\usepackage{footnote}
\newcommand{\source}{MAXI~J1810$-$222}
\newcommand{\sou}{J1810}

\newcommand{\swift}{\textit{Swift}}

\title[MAXI J1810--222 outflows]{An ultrafast outflow in the black hole candidate MAXI J1810-222?}

\author[M. Del Santo et al.]{
M. Del Santo$^{1}$\thanks{E-mail: melania.delsanto@inaf.it}, 
C. Pinto$^{1}$, A. Marino$^{2,3,1}$, A. D'A\`{i}$^{1}$, P.-O. Petrucci$^{4}$, J. Malzac$^{5}$,
J. Ferreira$^{4}$,  F. Pintore$^{1}$, 
\newauthor 
S.E. Motta$^{6}$,
T.D. Russell$^{1}$, A. Segreto$^{1}$, A. Sanna$^{7}$
\\
$^{1}$INAF, Istituto di Astrofisica Spaziale e Fisica Cosmica, Via U. La Malfa 153, I-90146 Palermo, Italy\\
$^{2}$Institute of Space Sciences (ICE, CSIC), Campus UAB, Carrer de Can Magrans s/n, E-08193 Barcelona, Spain\\
$^{3}$ Institut d'Estudis Espacials de Catalunya (IEEC), Carrer Gran Capit\`a 2--4, E-08034 Barcelona, Spain\\
$^{4}$ Univ. Grenoble Alpes, CNRS, IPAG, 38000 Grenoble, France\\
$^{5}$ IRAP, Université de Toulouse, CNRS, UPS, CNES, Toulouse, France \\
$^{6}$INAF, Osservatorio Astronomico di Brera, via E. Bianchi 46, 23807 Merate (LC), Italy\\
$^{7}$ Dipartimento di Fisica, Università degli Studi di Cagliari, SP Monserrato-Sestu km 0.7, 09042 Monserrato, Italy\\
}
\date{Accepted XXX. Received YYY; in original form ZZZ}

\pubyear{2023}

\begin{document}
\label{firstpage}
\pagerange{\pageref{firstpage}--\pageref{lastpage}}
\maketitle

\begin{abstract}
The transient X-ray source \source\ was discovered in 2018 and has been active ever since. A long combined radio and X-ray monitoring campaign was performed with ATCA and \swift, respectively.
It has been proposed that \source\ is a relatively distant black hole X-ray binary, albeit showing a very peculiar outburst behaviour.
Here, we report on the spectral study of this source making use of a large sample of NICER observations performed between 2019 February and 
2020 September.
We detected a strong spectral absorption feature at $\sim$1 keV,
which we have characterised with a physical photoionisation model.
Via a deep scan of the parameters space, we obtained evidence for a spectral-state dependent outflow, with mildly relativistic speeds.
In particular, the soft and intermediate states point to
a hot plasma outflowing at 0.05-0.15 $c$. 
This speeds rule-out thermal winds and,
hence, they suggest that such outflows could be radiation pressure
or (most likely) magnetically-driven winds. 
Our results are crucial to test current theoretical models of wind 
formation in X-ray binaries.

\end{abstract}




\section{Introduction}
Black hole X-ray transients (BHTs) spend most of their lifetimes in quiescence, 
which is characterised by an X-ray luminosity of $L_x < 10^{32}$ erg s$^{-1}$.
The active accretion phase, i.e. outburst, which can have a peak luminosity of $10^{36-39}$ erg s$^{-1}$,
can last from a few weeks up to several months \citep{tetarenko16}.
However, years-long outbursts have been observed from some sources, such as GRS 1915+105 \citep[see, e.g., ][]{motta21}.
During a typical outburst, BHTs usually show different 
spectral states, such as the canonical disc-dominated soft state 
and the Comptonisation dominated hard state, as well as a few intermediate states \citep[see, e.g., ][]{belloni16}.
Equatorial outflows in the form of disc winds are usually detected in disc-dominated states of sources seen at high inclination. This might depend on the wind radial density profile \citep[the closer our line-of-sight is to the disc
plane, the higher the optical depth, see Parra et al. sub.,][]{ponti12}.  
X-ray winds are mainly probed through resonant transitions of highly ionised elements
blue-shifted by Doppler motions with velocities of the order 
of few hundreds km/s. 
This indicates a lower limit on the escaping wind radius of the order of $10^{4-5} R_{\rm g}$. 
Here, the thermal motion of ions, due to irradiation of X-ray photons from the inner disk, is sufficient to unbind them from the gravitational well of the compact object (\textit{thermally-driven} winds, \citealt{Begelman1983}). 
However, such explanation is not universal.
In a few BH systems, such as GRO J1655-40, it has been suggested that \textit{magnetically-driven} winds provide a better description owing to their small launching radius, which may result in a higher escape velocity (\citealt{miller06}, but see \citealt{Tomaru2023}). 
The presence of a mildly-relativistic outflow in a sub-Eddington ($\dot{M} \lesssim 0.1 \dot{M}_{\rm Edd}$) source would be a strong evidence for a magnetic drive
(see, e.g., \citealt{chakra23, Fukumura2021, chakra16}). \\
\source\ (hereafter \sou) was discovered on 2018 November 29 by MAXI \citep{negoro18} and observed by {\it{NuSTAR}} during a soft spectral state.
No evidence of X-ray pulsations or type-I bursts in the light curve \citep{oeda19}. 
Around 2019 August, \swift/BAT survey data showed an increase in its hard X-ray emission, from which a radio/X-ray monitoring campaign with the Australia Telescope Compact Array (ATCA) and \swift/XRT was triggered.
The results of this 2-years long multi-wavelength (MW) campaign have been reported in \citet[hereafter R22]{russell22}.
From the X-ray spectral-timing properties and the radio behaviour, these authors argued that \sou\ is a relatively distant ($>$ 8 kpc) black hole candidate (BHC) with a very peculiar outburst behaviour. 
Indeed, unlike most BHTs, \sou\ was discovered in a soft state and does not follow the canonical hardness-intensity diagram (HID) since it goes back and forth from the hard to the soft state several times (across intermediate states; R22). 
At the time of writing this paper, a new X-ray/radio campaign is on-going, since, surprisingly (after about 4 years) \sou\ is still active
making it a BHC with years-long outburst.\\
In R22, we presented only three NICER observations representative of the different spectral states, in order to explore the overall variability properties of the source.
Here, we exploited a large sample of NICER observations to focus on narrow spectral features and their response to the continuum variations.

\begin{figure}
\centering
\includegraphics[width=0.7\columnwidth, angle=-90]{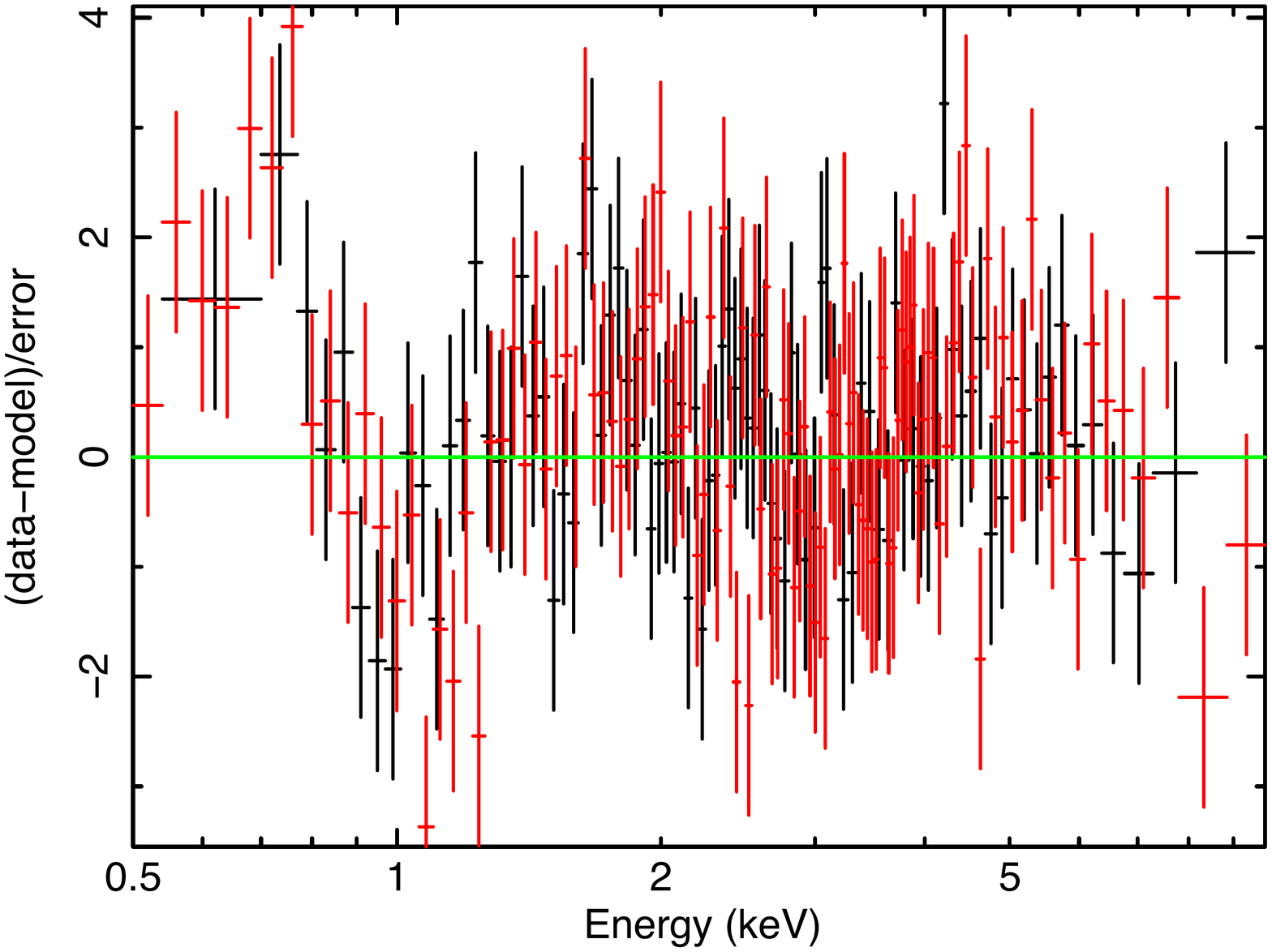}
\vspace{-0.5cm}
\caption{Residuals of the fit performed on the XRT spectra (in PC and WT modes) in the intermediate state.}
\label{fig:XRT}
\end{figure}


\section{Observations and data analysis}
\label{sec:ObsID}
We analysed all NICER public data collected from 2019 February 11 until 2020 September 25 (Target IDs 120056, 220056, 320056) for a total of 75 observations (ObsID). From this data-set, we visually inspected the light curves and excluded seven pointings (220056-0105/0106/0107/0108/0112, 2200560142, 3200560131) affected by background bright flares.
We reprocessed the NICER data using HEASoft version
6.29c and the NICER Data Analysis Software ({\textsc{nicerdas}})
version 8 with Calibration Database
({\textsc{caldb}}) version {\texttt{xti20210707}}. Spectra were extracted adopting standard calibration and screening criteria (the hot detectors 14 and 34 have been additionally excluded) using the  tool {\textsc{nicerl2}}. 
The background spectrum was obtained using the {\texttt{nibackgen3C50}} tool. 

We verified that our results are fully consistent with those
obtained by using the SCORPEON background and 
the most recent NICER calibrations (\texttt{xti20221001}) and pipeline (\textsc{nicerl3-spect}), 
that have been released during the submission phase of this manuscript.

\begin{figure}
\centering
\includegraphics[width=0.95\columnwidth]{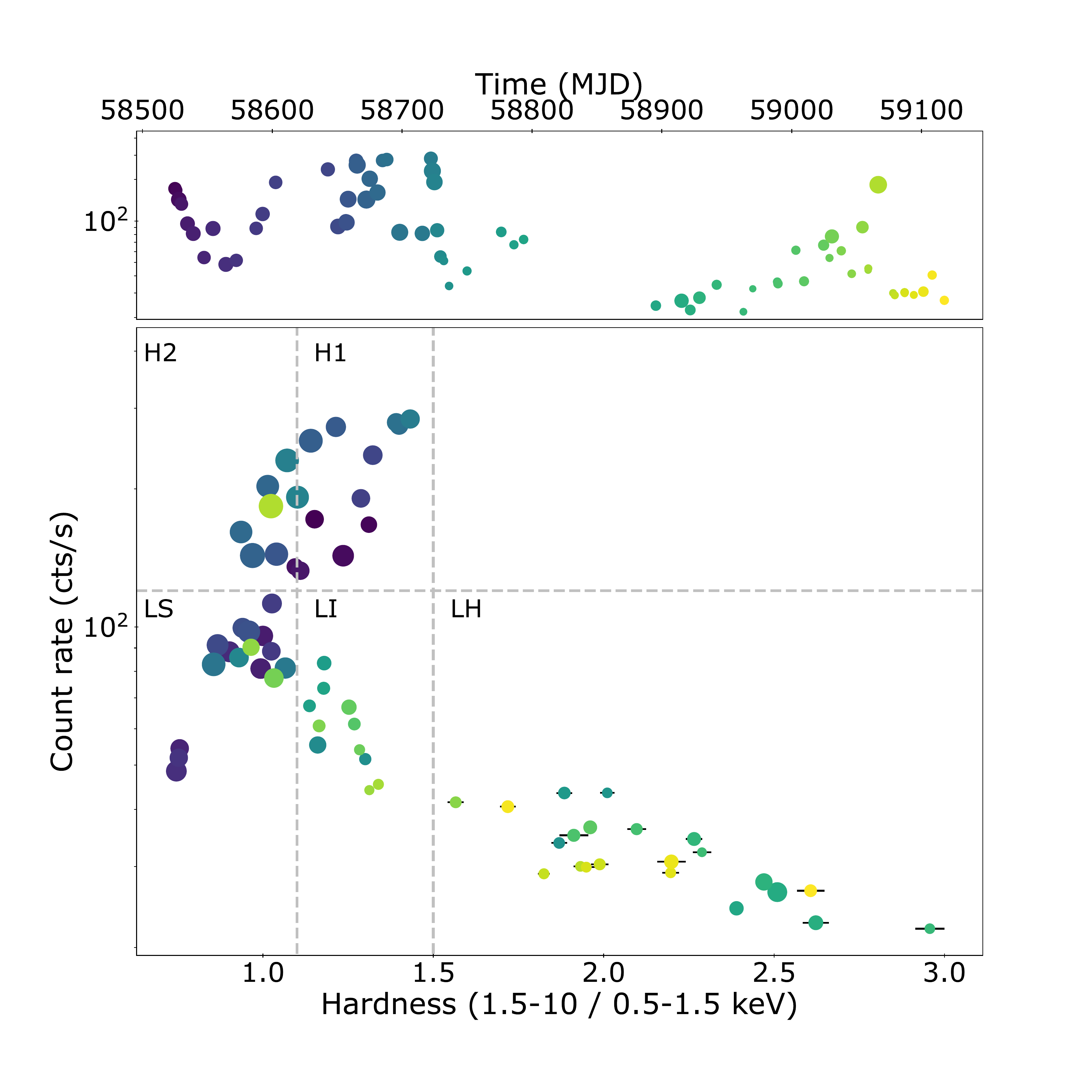}
\vspace{-0.5cm}
\caption{The NICER lightcurve (\textit{top}) and the HID (\textit{bottom}) 
of \sou\ are plotted (see Sec. \ref{sec:individual}). 
The time evolution is color coded from violet to yellow. 
The size of the points in the HID and light-curve refers to
the depth of the absorption line derived from the spectral fits.
The grey lines refer to the extraction of the five HID-resolved average spectra 
(see Fig. \ref{fig:5spec})
}.
\label{fig:HID}
\end{figure}

\subsection{Individual spectra}
\label{sec:individual}
First, we modelled the spectrum of each ObsID with an absorbed accretion disc 
plus power-law ({\texttt{tbabs(diskbb+po)}} in 
{\scriptsize{XSPEC}}). We applied a uniform 1\% systematic error to all spectra. 
We found a strong spectral feature at $\sim$1 keV in the residuals which can be well fitted with an absorption Gaussian line (\texttt{gabs} in {\scriptsize{XSPEC}}).
We noticed that the energy of the absorption line as well as the depth
are variable. 
In particular, the line energy spans (accounting for the uncertainties) from 0.84 keV to 1 keV, while the depth varies from 0.03 keV up to 0.21 keV.\\
We checked whether this feature could be instrumental, however, we notice that,
it has been also found in the \swift/XRT spectra (both in PC and WT modes) obtained by stacking all XRT data (presented in R22) in the intermediate state (see Fig. \ref{fig:XRT}). 
In Fig. \ref{fig:HID}, we show the long-term light-curve (top)
and the hardness-intensity diagram (HID; bottom) where each point corresponds to a single NICER ObsID.
The total background subtracted count rate is extracted in 0.5-10 keV, while the hardness is the ratio between the rates in the 1.5-10 keV and 0.5-1.5 keV bands.
It is worth noticing that the HID of \sou\ shows a peculiar pattern with respect to the 
q-track shape observed usually in BHTs, as also pointed out in R22.\\
The size of the points in the HID and light-curve (Fig. \ref{fig:HID}) refers to the depth of the absorption line derived from the spectral fits.
It can be noted that the depth value of the absorption line is connected to the spectral hardness: it is stronger in the soft spectra (LS) than in the hard ones (LH).
This value is (almost) the same in the spectra 
in the high flux branch (H1 and H2).

\subsection{Stacked spectra}
\label{sec:stacked}
An accurate characterisation of the line feature 
requires a strong detection and thereby, 
high signal-to-noise ratio (SNR) spectra.
We thus selected a total of five regions onto the HID from which 
we extracted the corresponding averaged spectra. 
The limits of the boundary regions (grey lines in Fig. \ref{fig:HID}, bottom)
were chosen in order to take into
account both the flux and the HR, as well as to ensure a sufficient count statistics per spectrum.
We named the regions with flux lower than 120 ct/s as
Low Soft (LS), Low Intermediate (LI), Low Hard (LH)
depending on their HR (see Tab. \ref{tab:hid}).
Then, the two regions with high flux (above 120 ct/s) have been called 
High 1 (H1) and High 2 (H2).

\begin{figure}
\centering
\vspace{-0.5cm}
\includegraphics[width=0.8\columnwidth, angle=-90]{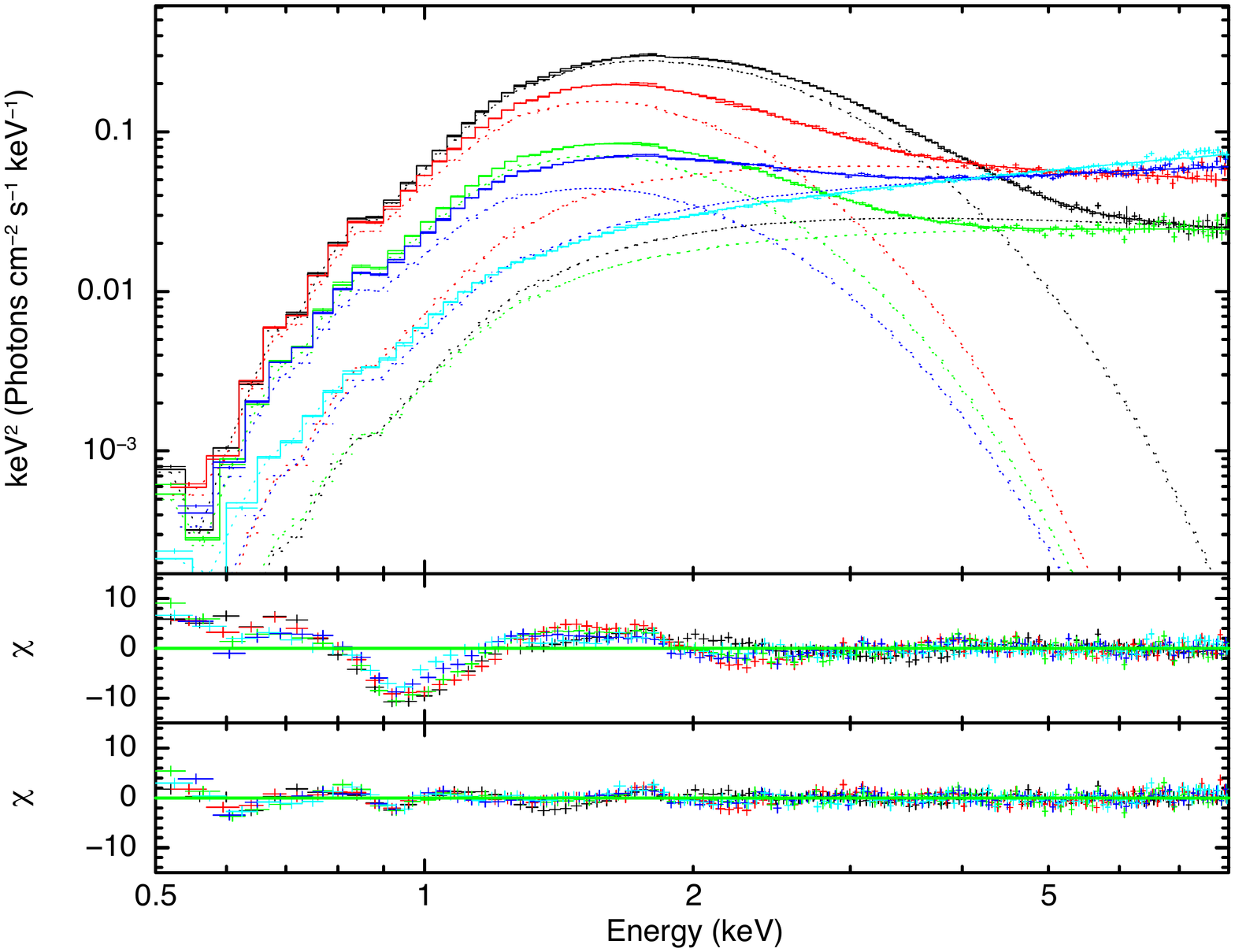}
\caption{Spectral continuum modelling of the five NICER HID-based stacked spectra  (see Fig.\,\ref{fig:HID}). From top to bottom progressively harder spectra are shown (top panel).
 In the middle panel, the residuals obtained from the continuum fitting are shown. 
 The energy centroid of the 1 keV feature decreases towards the hard state.
 Residuals of the best-fit continuum+gaussian model are show in the bottom panel. 
 }
\label{fig:5spec}
\end{figure}

\begin{table}
    \caption{HR and flux limits to select the five regions in the HID. 
     We report the best-fit parameters in last three column: the inner disc black-body temperature (kT$_{\rm in}$), energy and depth of the absorption line. The electron temperature of the Comptonisation component has been frozen at the best-fit values reported in R22 (see text).}
    \centering
    \begin{tabular}{c c c c c c }
    \hline
          State & HR & Flux & kT$_{\rm in}$ &  E$_{\rm line}$& Depth\\
          
          &  & (ct/s) & (keV) & (keV)  & (keV)\\
      \hline
      H1 & $>1.1$                   & $> 120$       & 0.49\,$\pm$\,0.01  & 0.96$\pm$\,0.01 & 0.13$\pm$\,0.02  \\
      H2 &  $\leq 1.1$              & $> 120$       & 0.37\,$\pm$\,0.01  & 0.97$\pm$\,0.02 & 0.15$\pm$\,0.02  \\
      LS & $\leq 1.1$               & $\leq 120$    & 0.38\,$\pm$\,0.01  & 0.95$\pm$\,0.01 & 0.12$\pm$\,0.02 \\
      LI & $ > 1.1$; $\leq 1.5$     &  $\leq 120$   & 0.39\,$\pm$\,0.01  & 0.93$\pm$\,0.01 & 0.06$\pm$\,0.01 \\
      LH & $>1.5$                   & $\leq 120$    & (0.2)  & 0.90$\pm$\,0.02 &  0.06$\pm$\,0.02\\
      
       \hline
    \end{tabular}
    \label{tab:hid}
\end{table}

We obtained five stacked spectra which have been modelled with a disc black-body emission plus a thermal Comptonisation component ({\texttt{nthcomp}} in {\scriptsize{XSPEC}},  Fig. \ref{fig:5spec}, top panel).
The electron temperature parameter ($kT_{\rm e}$) of the Comptonisation was frozen to the values found with the \swift/BAT hard X-ray spectra (R22), i.e., 15 keV in H1 and H2, 30 keV in LS and LI, 50 keV in LH. The seed photons temperatures were tied to the inner disc black-body temperature parameter (see Tab. \ref{tab:hid}).
The NICER spectra show a strong P-Cygni-like spectral feature around 1 keV which clearly decreases towards the hard spectrum
(see Fig. \ref{fig:5spec}, middle panel).
We also observed weaker features at 1.3-1.5 keV and at 2 keV.
We then included the {\texttt{gabs}} component to fit the 1 keV feature and obtain the best-fit parameters ($\chi^{2}_{\rm red}$(d.o.f.)=1.2(591); see Tab. \ref{tab:hid} and residuals in Fig. \ref{fig:5spec}, bottom panel).
Uncertainties are given at the 90\,\% confidence level.
In the hard spectrum, the temperature of the seed photons from the disc has been
frozen to 0.2 keV. It is worth noticing that, despite the HR values, H1 is the softest spectrum due to the fact that above 1.5 keV the disc contribution is still high 
with respect to the Comptonisation component.
The value for the neutral absorption column is always found consistent with the expected Galactic value ($\sim 0.9 \times$10$^{22}$ cm$^{-2}$).\\
Importantly, the 1 keV feature varies in centroid and in depth (see Tab. \ref{tab:hid}) and it is highly correlated with the spectral state (as we also find with a more physical model, see Sec. \ref{sec:model}). This most likely rules out any origin caused by a bad modelling of the interstellar medium (ISM). 
Moreover, the slow (daily) flux variation and the corresponding response
of the feature suggest that plasma is likely in photoionisation equilibrium.



\section{Photoionisation modeling}\label{sec:SED}
\begin{figure*}
\centering
\includegraphics[width=0.9\columnwidth]{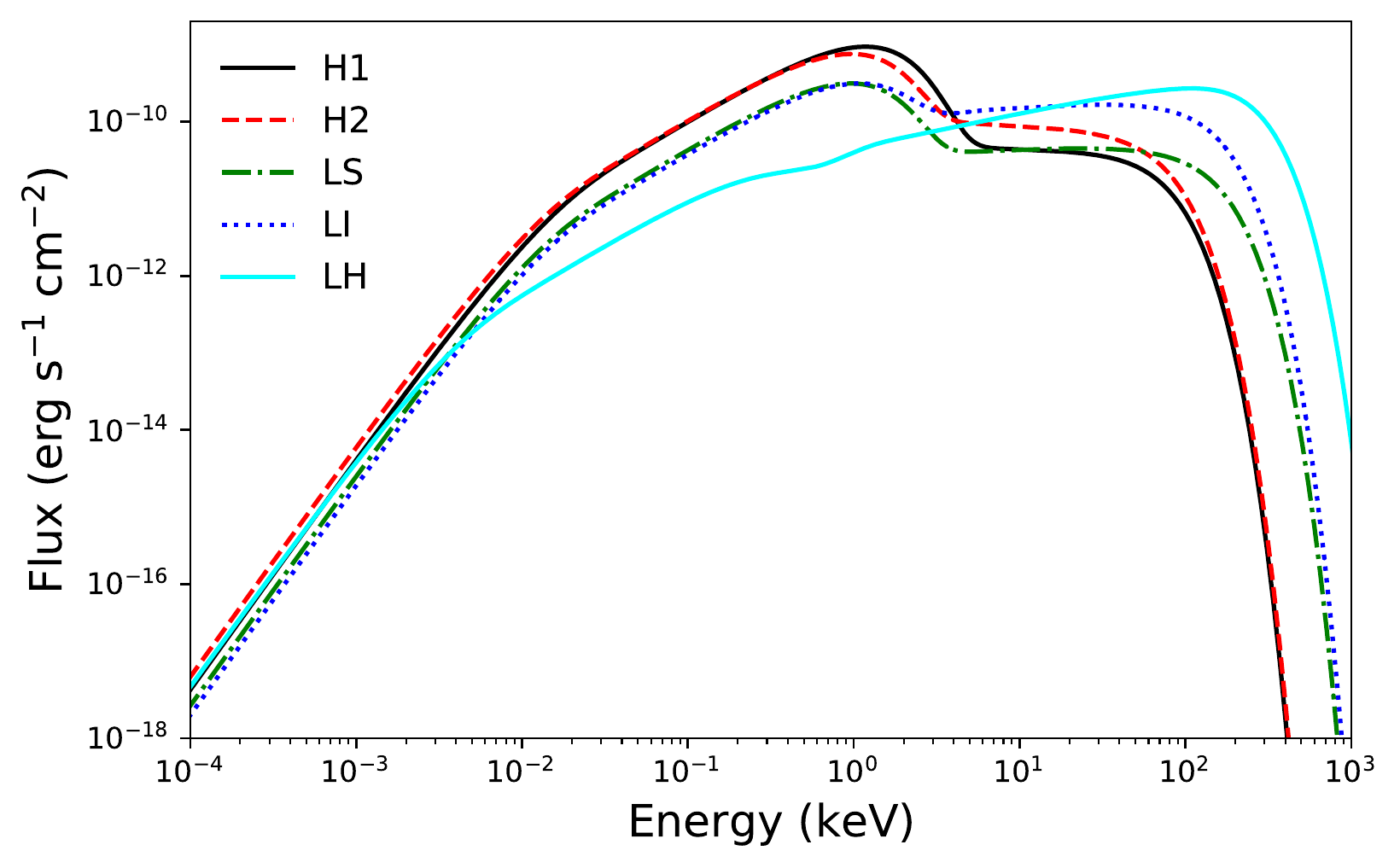}
\includegraphics[width=0.83\columnwidth]{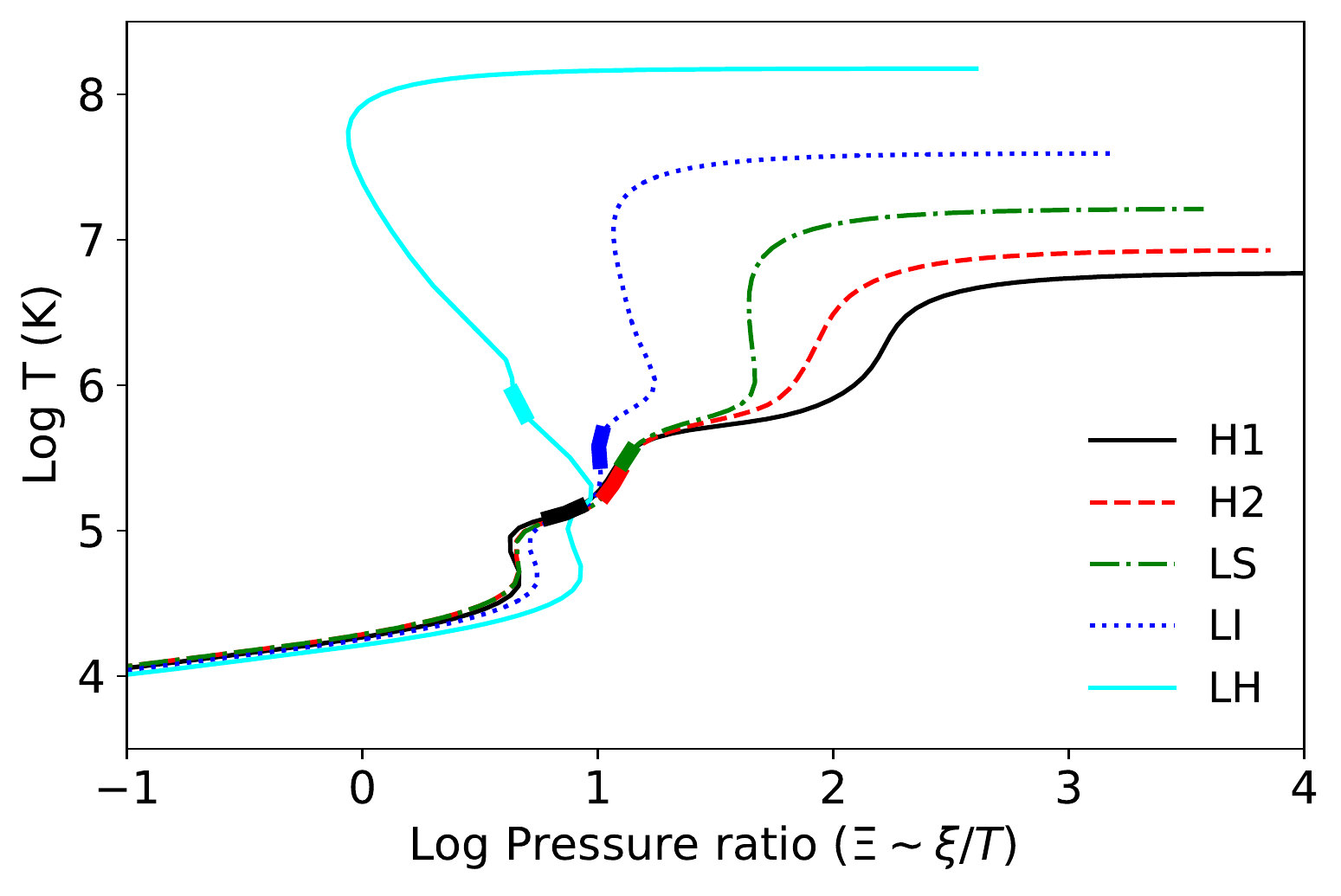}
\caption{
Left panel: SED for the five HID-resolved NICER stacked spectra. The absorption of the ISM was removed (intrinsic SEDs) to enable photoionisation balance computation. 
Righ panel: The stability curves of the five HID-resolved NICER spectra. Thicker segments 
show the (T-$\Xi$) ranges of the best-fit solutions.}
\label{fig:SED}
\end{figure*}

We performed a photoinisation modeling to derive physical parameters from the absorption feature. To this aim, we used {\scriptsize{SPEX}} which provides 
a faster photoionisation calculation ({\texttt{xabs}} component) compared to other codes (such as  {\scriptsize{XSTAR}} in {\scriptsize{XSPEC}}).\\
Once we obtained the best-fit continuum model for each of the five stacked spectra  with similar models as in {\scriptsize{XSPEC}}, i.e., {\texttt{dbb+comp}} in {\scriptsize{SPEX}}, 
we extrapolated it down between the optical band (10$^{-4}$ keV) and  the hard X-rays (100 keV), to build a Spectral Energy Distribution (SED) broad enough (see Fig. \ref{fig:SED}, left panel) to compute the photoionisation balance with the task {\scriptsize{XABSINPUT}}. 

This photoionisation balance links the gas temperature ($kT$)
and the ionisation state of the plasma characterised by the ionisation parameter $\xi = L / n_{\rm H} R^{2}$,
where $L$ is the ionising luminosity of the source, $n_{\rm H}$ is the plasma volume density, and $R$ is its distance from the ionising source. 
We also computed the stability curves for each SED, which show the relationship between the temperature and the pressure ratio (see Fig. \ref{fig:SED}, right panel). The latter is defined as the ratio between the radiation and the thermal pressure which results in $\Xi = 19222 \, \xi / T$ \citep{Krolik1981}. These curves are very useful because at positive (negative) slopes the irradiated plasma is in a stable (unstable) solution (see, however, \citealt{pop21} regarding the potential caveats when computing these stability curves). 

\subsection{Spectral fitting}
\begin{figure*}
\centering
\includegraphics[width=0.67\columnwidth]{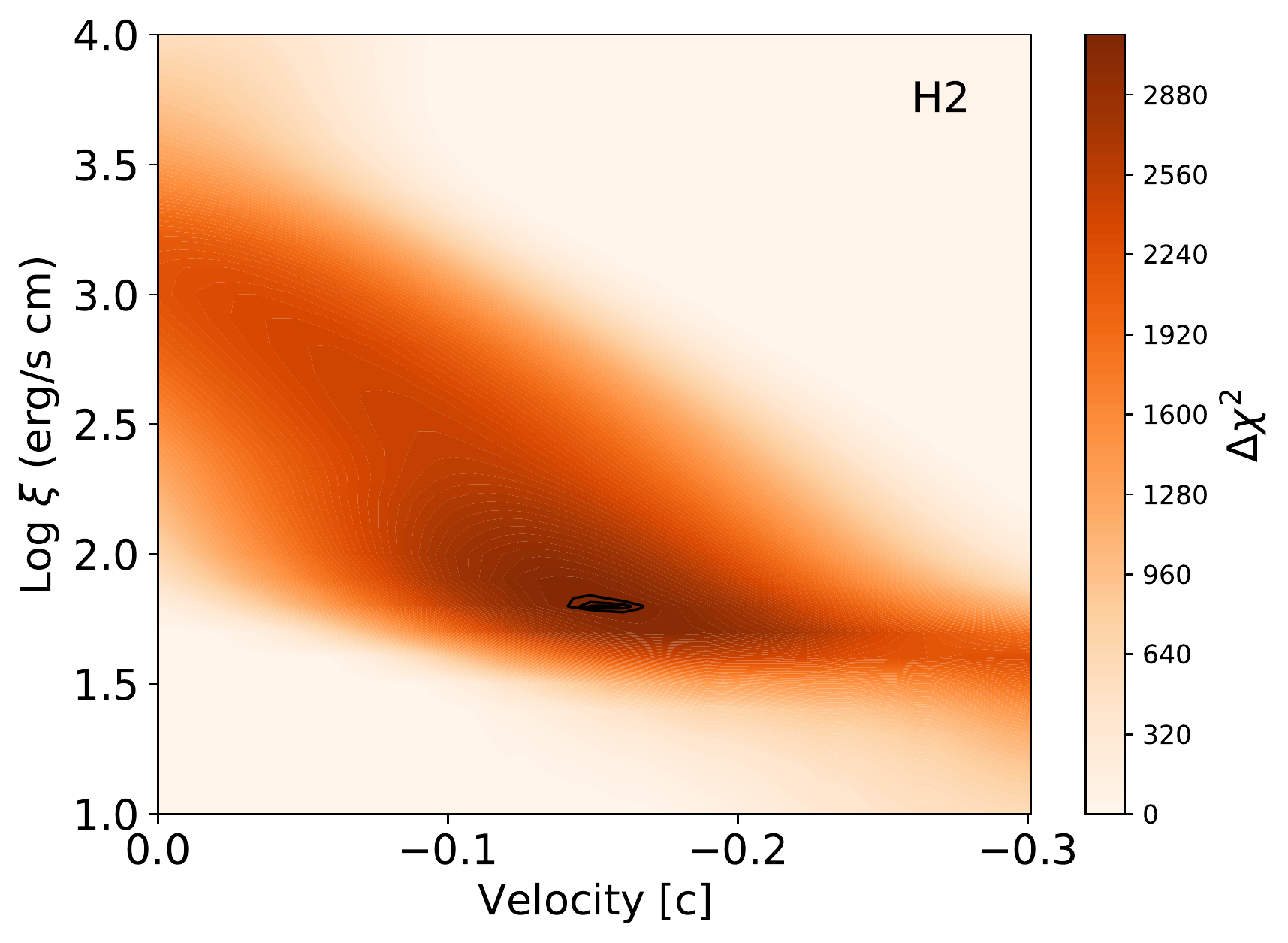}
\includegraphics[width=0.67\columnwidth]{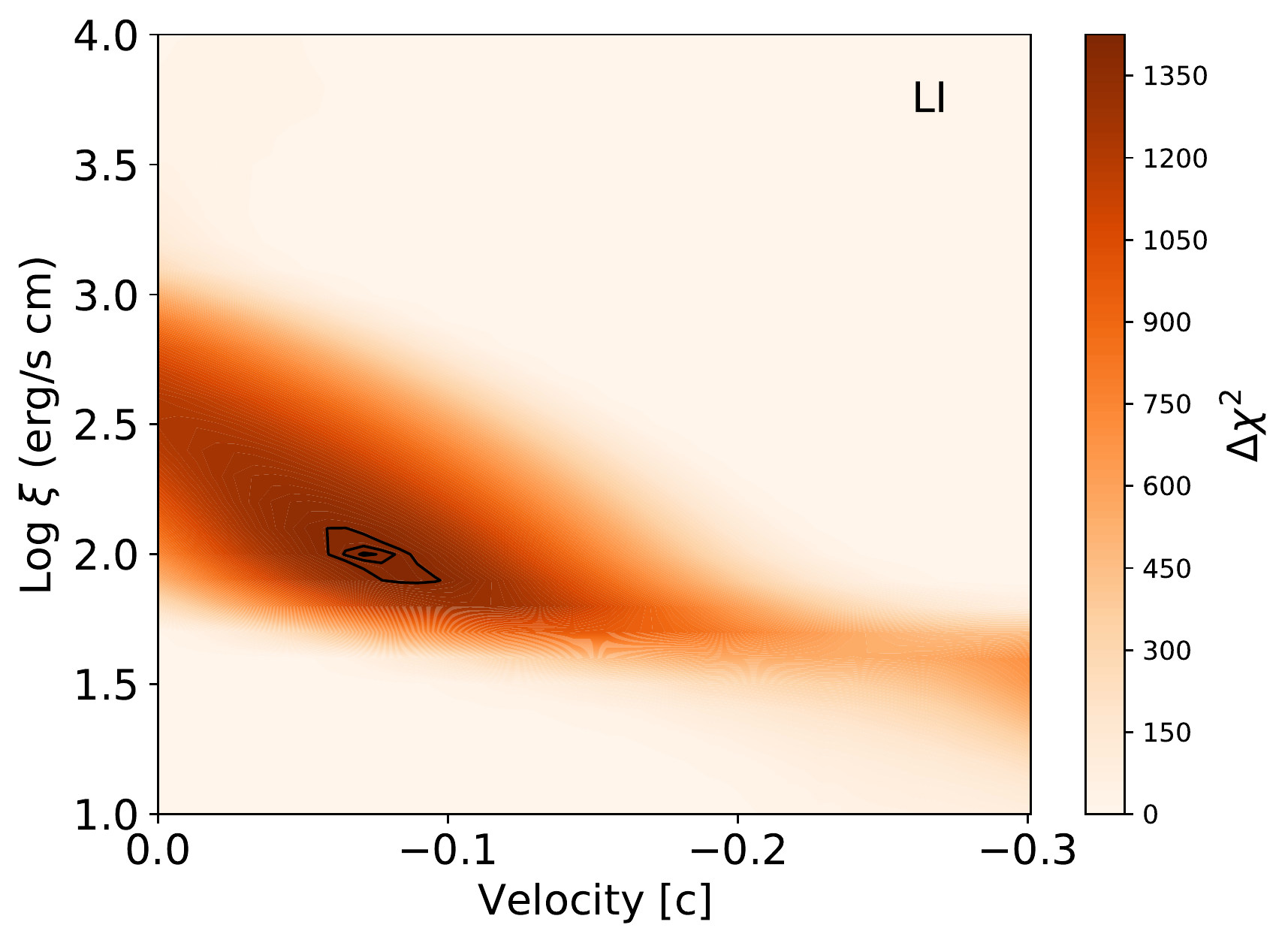}
\includegraphics[width=0.67\columnwidth]{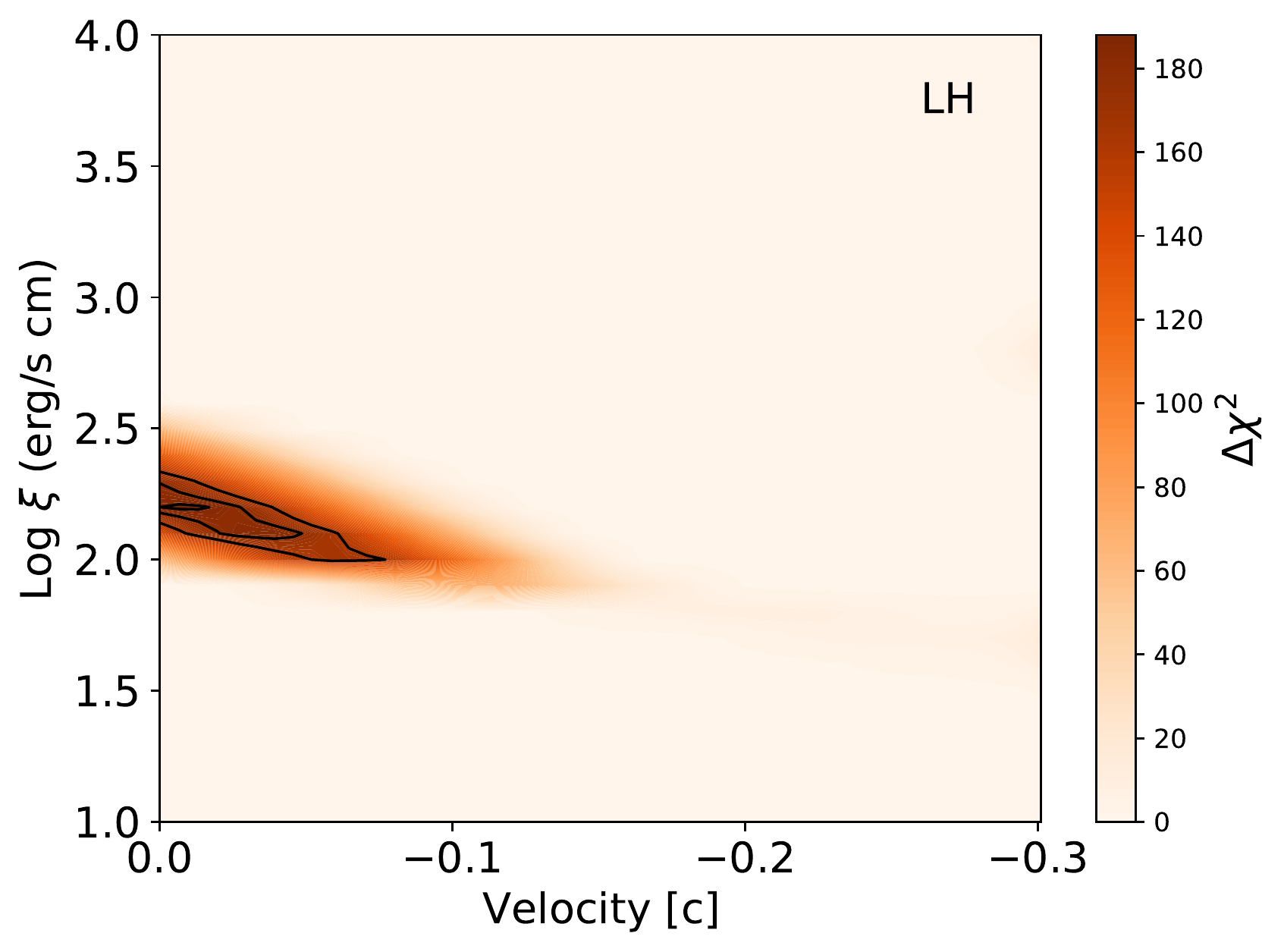}
\vspace{-0.25cm}
\caption{Results from the multi-dimensional scan with the model grids for the spectra from the H2, LI and LH epochs (left to right, respectively). Reported are the {\texttt{xabs}} improvements to the continuum-only fits. The three black solid lines indicate 1, 2 and 3$\sigma$ contours for 2 free parameters ($\xi,v_{\rm LOS}$).}
\label{fig:grids}
\end{figure*}

The output file produced by the {\scriptsize{XABSINPUT}} task is used by the  {\texttt{xabs}} model to make a list of absorption lines for each element ion and fit the spectrum. This allowed us to characterise the absorption features. The main parameters of {\texttt{xabs}} are: the column density, $N_{\rm H}$, ionisation parameter, $\xi$, line-of-sight velocity $V_{\rm LOS}$, and velocity dispersion or line width, $V_{\sigma}$. 
Given the low spectral resolution, we assumed Solar chemical abundances.\\
To avoid getting stuck into local minima of the $\chi^2$ distribution, 
we performed a multi-dimensional scan in the $V_{\rm LOS}$ -- log $\xi$ parameter space \citep{pinto21}. Fig. \ref{fig:grids} shows the results for the H2, LI and LH NICER spectra assuming a velocity dispersion of $V_{\sigma}=$10,000 km/s. These are consistent with those obtained with 100 and 1,000 km/s because at 1 keV NICER lacks the spectral resolution necessary to resolve individual lines. The colour in Fig. \ref{fig:grids} is coded according to the $\chi^2$ improvement on top of the spectral continuum ($\Delta \chi^2$). The velocities are negative, i.e., blueshifts corresponding to outflows. 
The $\Delta \chi^2$ are large indicating a very high significance, well above 5 $\sigma$, even accounting for the look-elsewhere effect implying that there is no need to perform dedicated Monte Carlo simulations.

\begin{figure}
\centering
\includegraphics[width=0.9\columnwidth, angle=0]{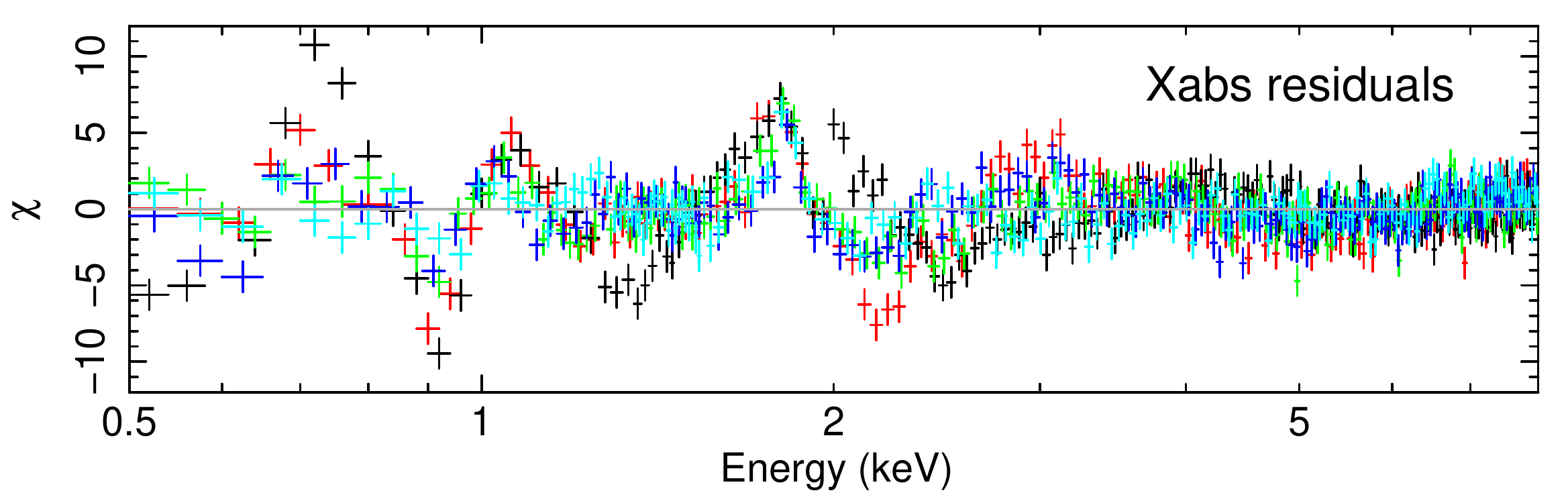}
\caption{Residuals for the spectral fits with the continuum + photoionised absorber model for the five NICER stacked spectra (see also  Fig.\,\ref{fig:HID} and \ref{fig:5spec}).}
\label{fig:xabs_res}
\end{figure}

Thereafter, to determine the plasma parameters, we refit the five averaged spectra with the \texttt{xabs} model starting from the best-fit solutions obtained by the grids. The residuals around 1 keV have been significantly decreased including those at other energies implying that multiple features are described
(see Fig. \ref{fig:xabs_res}). 
Further minor residuals are left likely due to a more complex plasma structure or the presence of rest-frame emission lines (i.e., a P-Cygni profile). Moreover, the stacking of spectra from the same HID region may have also broadened the absorption measure distribution of the photoionised absorber. In Table \ref{tab:results} we show the main \texttt{xabs} parameters from the best fits.


\section{Results and discussion}\label{sec:model}
\begin{figure}
\centering
\includegraphics[width=0.8\columnwidth]{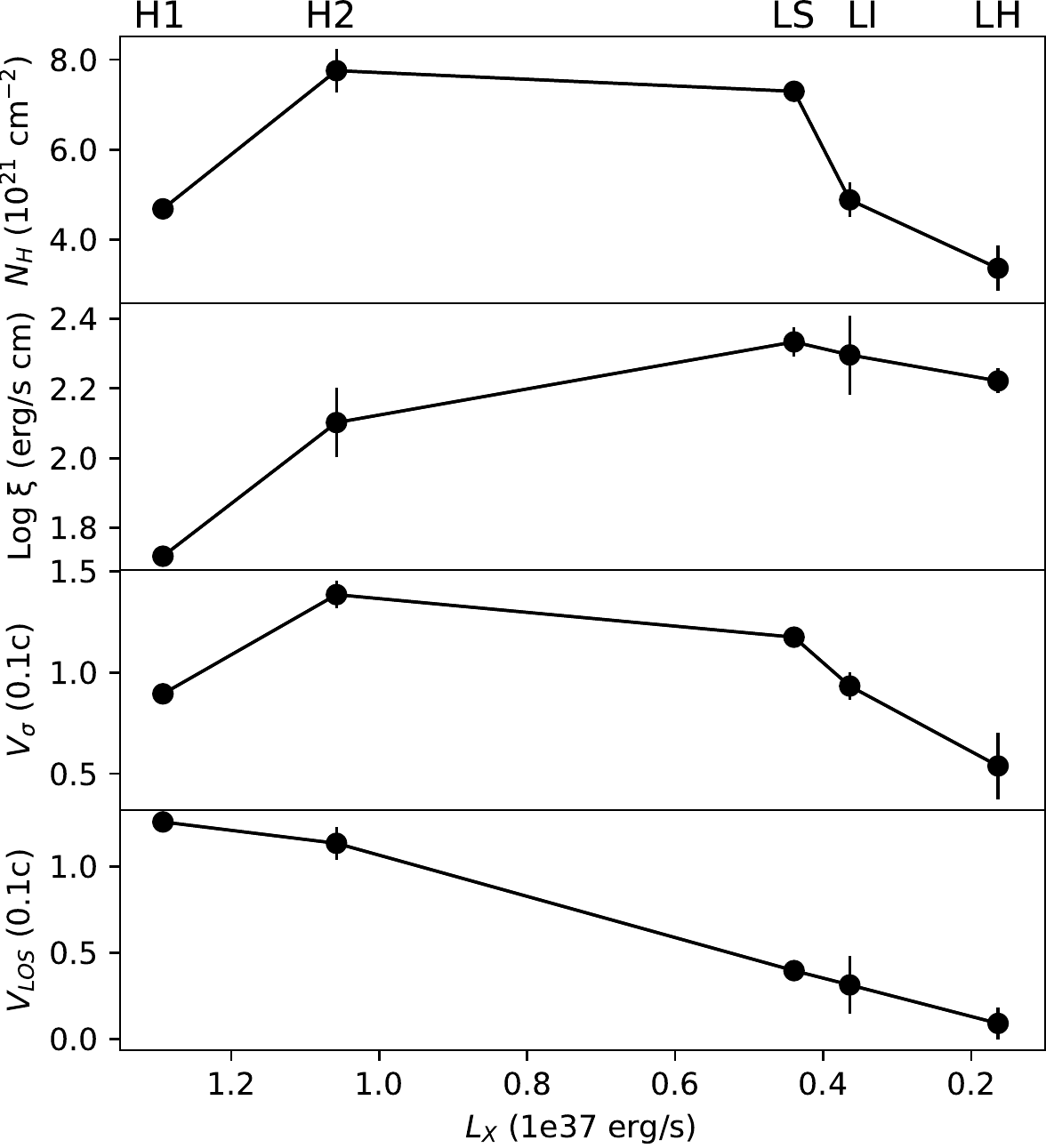}
\vspace{-0.25cm}
\caption{Best-fit parameters of the outflowing plasma component ({\texttt{xabs}}) for the five NICER spectra sorted according to the X-ray 0.3-10 keV luminosity. The X-axis is inverted following the HID evolution from high-soft to low-hard states (see Fig. \ref{fig:HID}).}
\label{fig:resuls}
\end{figure}

We have found a strong feature in the NICER spectra of \sou\ which we interpret as a blend of blue-shifted Fe L, Ne X or O VIII absorption lines. This is confirmed by the ionic column densities yielded by {\texttt{xabs}} for log $\xi \sim 2$.
In Fig. \ref{fig:resuls}, we show the trend of the main best-fit {\texttt{xabs}} parameters with the source spectral state, as a function of the X-ray luminosity. 
In agreement with the results of the empirical line analysis (see Sect. \ref{sec:stacked}), the plasma state significantly varies with the source spectral state. The high flux states are characterised by a 
highly-significant solution which indicates a hot plasma outflowing at $>0.1c$, which is unlikely for classical Galactic X-ray binary (XRB) thermal winds ($v_{\rm XRB}<1000$ km/s, i.e. $\sim 0.0033c$, see \citealt{tomaru19}) and must involve either strong radiation pressure (as in ULXs winds, see \citealt{pinto16}) or (most likely) magnetically driven winds. 
However, a ULX nature of \sou\ is unlikely given the hard spectra observed.
In addition, a ULX nature would require a distance of about 100 kpc which is
outside the Galaxy.\\
In the LS and LI states the derived velocities, albeit lower, are still extreme for a thermal wind.
However, in the hard state the outflow is much weaker, due to a low $N_{\rm H}$, and slower, similarly to thermal winds in XRBs.\\
The changes of the wind properties between the different spectral states of the source might be related to the thermal instability of the plasma  since 
the ($T,\xi,\Xi$) solution for the hard state is on the edge between the stable and unstable branches with $1<\log \xi <3$, corresponding to temperatures $10^5 < T < 10^7$ K (see Fig. \ref{fig:SED}, right panel).
It is worth noticing that the LOS velocity (if associated with the terminal velocity) should be higher than the velocity dispersion,
which is not always the case in our results. 
This is likely due to the lack of spectral resolution of NICER or 
the spectral stacking and, therefore, the absolute values should be taken with caution.\\
On the other hand, the variability of the wind parameters may be due to a different configuration of the magnetic field and launching radius. For instance, if we assume that the outflow speed equals the escape velocity ($v_{\rm esc}=\sqrt{2GM/R}$), we estimate that all, but one, spectra
 have launching radius between 100 and 2000 $R_{\rm g}$ (progressively
 increasing from the H1 to the LI state). The large value of 24000 $R_{\rm g}$ derived in the LH state might be an over-estimation due to the line weakness (see contours in Fig. \ref{fig:grids}) 
 although it would be close to the expected range of values for a thermally-driven wind \citep{Begelman1983}.
From the definition of the ionisation parameter we estimate density decreasing from $8 \times 10^{18}$ to $3\times10^{13}$ cm$^{-3}$, which are densities that range from XRB inner disc to typical XRB outflows \citep{Bianchi2017}.
These are likely upper limits on the density as the outflow may be launched at larger radii and accelerated to the observed velocities by the magnetic fields.
Upper limits on the outflow distance can be derived by
considering the wind thickness, $\Delta\,R\leq R$, and the ionisation parameter definition,
$\xi = L / n_H R^2 = L / (n_H R \Delta\,R) \cdot \Delta\,R / R < L / (n_H R \Delta\,R) = L / (N_H R) \rightarrow R < L / (\xi N_H)$.
Upper limits ranging from a few $10^{6}$ R$_{\rm G}$ to $10^{7}$ R$_{\rm G}$ are obtained.\\
The launching radius, as estimated from the assumption of escape velocity, increases towards LH, following the evolution of the innermost radius of the optically thick Shakura-Sunyev disc (that reaches almost 100 $R_g$ in the hard state, see, e.g., \citealt{done07} and refs therein).
Alternatively, the decrease of both outflow velocity and column density in the hard state may be due to an increase in the collimation of the outflow, thus becoming a jet.\\
The outflow rate is defined as $\dot{M}_w = 4 \, \pi \, R^2 \, \rho \, v_w \, \Omega \, C$, where $\Omega$ and $C$ are the solid angle and the volume filling factor (or \textit{clumpiness}), respectively, $\rho=n_{\rm H}  \, m_p \, \mu$ is the gas density and $R$ is the distance from the ionising source. 
In the two high flux spectra, the outflow rate appears mildly super-Eddington (2-7 $\dot{\rm M}_{\rm Edd}$, assuming M$_{\rm BH}$ = 10 $\rm M_{\odot}$, efficiency 10\,\% and $C=0.1$). This may remove most of the material, perhaps explaining the transition to the low flux states where both the accretion and outflow rates are much smaller (0.1-0.2 $\dot{\rm M}_{\rm Edd}$). Such a value would still require an accretion rate of $0.1 \, \dot{\rm M}_{\rm Edd}$, which would correspond to an intrinsic luminosity of $10^{38}$ erg/s for a BH, suggesting that the source is even more distant,
i.e. at 20-30 kpc.
Although slightly high, our estimates of $\dot{M}_w$ are comparable with results from magneto-hydrodynamic (MHD) calculation of magnetically driven winds in other BHTs by, e.g., \cite{Fukumura2021}.\\
If we assume both solid angle and covering fraction of about 0.1, which are typical for extreme outflows (e.g., \citealt{pinto21}, although the latter can be as low as 0.01, see, e.g., \citealt{kobayashi18}), we estimate an outflow kinetic power ($L_w = 1/2 \, \dot{M}_w \, v^2$) between $10^{38}$ erg s$^{-1}$ (H1) and $10^{34}$ erg s$^{-1}$ (LH). This corresponds to, approximately,
10 to 0.01 L$_{\rm Bol}$ (Tab. \ref{tab:results}), 
which means that the outflow dominates the energetic budget in the soft state (both at high and low flux) and it decreases towards the hard state (L$_{w}$ $\sim$ 1\%).
Noteworthy, such an evolution is qualitatively consistent with the JED-SAD framework (\citealt{ferreira06, pop10, marcel22} and refs therein). While transiting from the soft to the hard state,  \sou\ would undergo a switch between an optically thick and geometrically thin inner disk launching winds that carry away a significant portion of the released accretion power, to an optically thin and geometrically thick disk where most of the accretion power would be actually advected into the black hole and powering jets, lowering significantly the bolometric luminosity. 
In MHD outflows, the flow speed at the Alfven point is comparable
to the rotation speed at the disk midplane (see, e.g., \citep{ferreira97}). 
For a wind speed of 0.1$c$, this provides an anchoring radius 
R$\sim$100 R$_{\rm g}$, with a required B-field of at least 10$^5$ G. 
A Jet Emitting Disk (JED, i.e., magnetic corona) is the accretion mode with the highest possible magnetic field value, 
where the dominant torque (allowing accretion) is that due to the bipolar jets. 
At R$\sim$100 R$_{\rm g}$ and for a BH mass of 10 M$_{\odot}$ a JED would have
B = 4 $\times 10^{5} (\dot{m})^{1/2}$ G, 
where $\dot{m} = \dot{M} c^{2}/L_{Edd}$ is the normalized disk accretion rate at that same radius \citep{pop10}.
Considering that we do not have a firm estimation of the distance,
this would result in a B field intensity of $10^{4-5}$ Gauss for \sou.



Finally, based on the high outflow speeds (between 0.03$c$ and 0.05$c$)
derived by lines detection, magnetically driven winds have been claimed in other BHTs, e.g., IGR J17091-3624 \citep{king12}, 4U 1630-472 \citep{king14} and  GRS 1915+105 \citep{miller16}. 
The latter and \sou\ show both a years-long outburst, while the other two 
show recurrent and frequent outbursts, thus indicating that the secular mass
accretion rate should be higher in these systems when compared to a typical BHT. 
However, the maximum value of the wind velocity in \sou\ is
about three times faster (i.e., 0.1$c$) indicating a more efficient launching mechanism.
An in-depth theoretical work combined with an observational study of a larger sample 
of ultra fast outflows in XRBs would place better constraints on the outflow mechanism.

\begin{table}
    \caption{Results from best-fit photoionisation model. The unabsorbed X-ray and Bolometric luminosities are computed between 0.3-10 keV and 0.001-1000 keV. {\texttt{Xabs}} parameters are the column density, $N_{\rm H}$, ionisation parameter, $\xi$, line-of-sight velocity $V_{\rm LOS}$, and velocity dispersion or line width, $V_{\sigma}$.}
\renewcommand{\arraystretch}{1.0}
 \small\addtolength{\tabcolsep}{-2pt}
\scalebox{0.925}{%
    \begin{tabular}{c c c c c c c}
    \hline
          State & L$_{\rm X}^a$ & L$_{\rm Bol}^a$ & $N_{\rm H}$ & Log ${\xi}$ & $v_{\sigma}$ & $v_{\,\rm LOS}$ \\
                & \multicolumn{2}{c}{$10^{37}$ erg/s} & $10^{21}$ cm$^{-2}$ & erg/s cm & $0.1c$  & $0.1c$ \\
      \hline
      H1 & 1.29 & 1.56 & $4.68 \pm 0.05$ & $1.72 \pm 0.01$ & $0.90 \pm 0.04$ & $1.26 \pm 0.01$ \\
      H2 & 1.06 & 1.4 & $7.75 \pm 0.49$ & $2.10 \pm 0.10$ & $1.39 \pm 0.07$ & $1.13 \pm 0.09$ \\
      LS & 0.44 & 0.6 & $7.29 \pm 0.23$ & $2.33 \pm 0.04$ & $1.18 \pm 0.04$ & $0.40 \pm 0.03$ \\
      LI & 0.36 & 0.6 & $4.88 \pm 0.39$ & $2.30 \pm 0.11$ & $0.93 \pm 0.07$ & $0.31 \pm 0.17$ \\
      LH & 0.16 & 0.7 & $3.36 \pm 0.51$ & $2.22 \pm 0.04$ & $0.54 \pm 0.16$ & $0.09 \pm 0.09$ \\
      
       \hline
    \end{tabular}}
Note: ($^a$) luminosities are reported without uncertainty given that the statistical error (a few \%) is much smaller than that on the distance (adopted at 8 kpc).  
    \label{tab:results}
\end{table}

\section*{Acknowledgements}
MDS, CP, ADA, SEM, FP, TDR acknowledge support from the INAF grant "ACE-BANANA".
CP acknowledges support from AHEAD2020 project (grant agreement n. 871158).
AM is supported by the H2020 ERC Consolidator Grant “MAGNESIA” (PI: Rea) and National Spanish grant PGC2018-095512-BI00.
AD acknowledges funding from the
contract ASI/INAF n. I/004/11/4. 
POP and JF acknowledge financial support from the french national high energy programme from INSU/CNRS and from CNES.

\section*{Data Availability}

All data used in this work are publicly available on the HEASARC NICER archive.
The codes for the multi-dimensional scan and the production of photoionisation model grids are publicly available (https://github.com/ciropinto1982).



\bibliographystyle{mnras}
\bibliography{biblio} 

\begin{thebibliography}{}
\makeatletter
\relax
\def\mn@urlcharsother{\let\do\@makeother \do\$\do\&\do\#\do\^\do\_\do\%\do\~}
\def\mn@doi{\begingroup\mn@urlcharsother \@ifnextchar [ {\mn@doi@}
  {\mn@doi@[]}}
\def\mn@doi@[#1]#2{\def\@tempa{#1}\ifx\@tempa\@empty \href
  {http://dx.doi.org/#2} {doi:#2}\else \href {http://dx.doi.org/#2} {#1}\fi
  \endgroup}
\def\mn@eprint#1#2{\mn@eprint@#1:#2::\@nil}
\def\mn@eprint@arXiv#1{\href {http://arxiv.org/abs/#1} {{\tt arXiv:#1}}}
\def\mn@eprint@dblp#1{\href {http://dblp.uni-trier.de/rec/bibtex/#1.xml}
  {dblp:#1}}
\def\mn@eprint@#1:#2:#3:#4\@nil{\def\@tempa {#1}\def\@tempb {#2}\def\@tempc
  {#3}\ifx \@tempc \@empty \let \@tempc \@tempb \let \@tempb \@tempa \fi \ifx
  \@tempb \@empty \def\@tempb {arXiv}\fi \@ifundefined
  {mn@eprint@\@tempb}{\@tempb:\@tempc}{\expandafter \expandafter \csname
  mn@eprint@\@tempb\endcsname \expandafter{\@tempc}}}

\bibitem[\protect\citeauthoryear{{Begelman}, {McKee}  \& {Shields}}{{Begelman}
  et~al.}{1983}]{Begelman1983}
{Begelman} M.~C.,  {McKee} C.~F.,   {Shields} G.~A.,  1983, \mn@doi [\apj]
  {10.1086/161178}, \href
  {https://ui.adsabs.harvard.edu/abs/1983ApJ...271...70B} {271, 70}

\bibitem[\protect\citeauthoryear{{Belloni} \& {Motta}}{{Belloni} \&
  {Motta}}{2016}]{belloni16}
{Belloni} T.~M.,  {Motta} S.~E.,  2016, in {Bambi} C.,  ed.,  Astrophysics and
  Space Science Library Vol. 440, Astrophysics of Black Holes: From Fundamental
  Aspects to Latest Developments. p.~61 (\mn@eprint {arXiv} {1603.07872}),
  \mn@doi{10.1007/978-3-662-52859-4_2}

\bibitem[\protect\citeauthoryear{{Bianchi}, {Ponti}, {Mu{\~n}oz-Darias}  \&
  {Petrucci}}{{Bianchi} et~al.}{2017}]{Bianchi2017}
{Bianchi} S.,  {Ponti} G.,  {Mu{\~n}oz-Darias} T.,   {Petrucci} P.-O.,  2017,
  \mn@doi [\mnras] {10.1093/mnras/stx2173}, \href
  {https://ui.adsabs.harvard.edu/abs/2017MNRAS.472.2454B} {472, 2454}

\bibitem[\protect\citeauthoryear{{Chakravorty} et~al.,}{{Chakravorty}
  et~al.}{2016}]{chakra16}
{Chakravorty} S.,  et~al., 2016, \mn@doi [\aap] {10.1051/0004-6361/201527163},
  \href {https://ui.adsabs.harvard.edu/abs/2016A&A...589A.119C} {589, A119}

\bibitem[\protect\citeauthoryear{{Chakravorty} et~al.,}{{Chakravorty}
  et~al.}{2023}]{chakra23}
{Chakravorty} S.,  et~al., 2023, \mn@doi [\mnras] {10.1093/mnras/stac2835},
  \href {https://ui.adsabs.harvard.edu/abs/2023MNRAS.518.1335C} {518, 1335}

\bibitem[\protect\citeauthoryear{{Done}, {Gierli{\'n}ski}  \& {Kubota}}{{Done}
  et~al.}{2007}]{done07}
{Done} C.,  {Gierli{\'n}ski} M.,   {Kubota} A.,  2007, \mn@doi [\aapr]
  {10.1007/s00159-007-0006-1}, \href
  {https://ui.adsabs.harvard.edu/abs/2007A&ARv..15....1D} {15, 1}

\bibitem[\protect\citeauthoryear{{Ferreira}}{{Ferreira}}{1997}]{ferreira97}
{Ferreira} J.,  1997, \mn@doi [\aap] {10.48550/arXiv.astro-ph/9607057}, \href
  {https://ui.adsabs.harvard.edu/abs/1997A&A...319..340F} {319, 340}

\bibitem[\protect\citeauthoryear{{Ferreira}, {Petrucci}, {Henri}, {Saug{\'e}}
  \& {Pelletier}}{{Ferreira} et~al.}{2006}]{ferreira06}
{Ferreira} J.,  {Petrucci} P.~O.,  {Henri} G.,  {Saug{\'e}} L.,   {Pelletier}
  G.,  2006, \mn@doi [\aap] {10.1051/0004-6361:20052689}, \href
  {https://ui.adsabs.harvard.edu/abs/2006A&A...447..813F} {447, 813}

\bibitem[\protect\citeauthoryear{{Fukumura}, {Kazanas}, {Shrader}, {Tombesi},
  {Kalapotharakos}  \& {Behar}}{{Fukumura} et~al.}{2021}]{Fukumura2021}
{Fukumura} K.,  {Kazanas} D.,  {Shrader} C.,  {Tombesi} F.,  {Kalapotharakos}
  C.,   {Behar} E.,  2021, \mn@doi [\apj] {10.3847/1538-4357/abedaf}, \href
  {https://ui.adsabs.harvard.edu/abs/2021ApJ...912...86F} {912, 86}

\bibitem[\protect\citeauthoryear{{King} et~al.,}{{King} et~al.}{2012}]{king12}
{King} A.~L.,  et~al., 2012, \mn@doi [\apjl] {10.1088/2041-8205/746/2/L20},
  \href {https://ui.adsabs.harvard.edu/abs/2012ApJ...746L..20K} {746, L20}

\bibitem[\protect\citeauthoryear{{King} et~al.,}{{King} et~al.}{2014}]{king14}
{King} A.~L.,  et~al., 2014, \mn@doi [\apjl] {10.1088/2041-8205/784/1/L2},
  \href {https://ui.adsabs.harvard.edu/abs/2014ApJ...784L...2K} {784, L2}

\bibitem[\protect\citeauthoryear{{Kobayashi}, {Ohsuga}, {Takahashi},
  {Kawashima}, {Asahina}, {Takeuchi}  \& {Mineshige}}{{Kobayashi}
  et~al.}{2018}]{kobayashi18}
{Kobayashi} H.,  {Ohsuga} K.,  {Takahashi} H.~R.,  {Kawashima} T.,  {Asahina}
  Y.,  {Takeuchi} S.,   {Mineshige} S.,  2018, \mn@doi [\pasj]
  {10.1093/pasj/psx157}, \href
  {https://ui.adsabs.harvard.edu/abs/2018PASJ...70...22K} {70, 22}

\bibitem[\protect\citeauthoryear{{Krolik}, {McKee}  \& {Tarter}}{{Krolik}
  et~al.}{1981}]{Krolik1981}
{Krolik} J.~H.,  {McKee} C.~F.,   {Tarter} C.~B.,  1981, \mn@doi [\apj]
  {10.1086/159303}, \href
  {https://ui.adsabs.harvard.edu/abs/1981ApJ...249..422K} {249, 422}

\bibitem[\protect\citeauthoryear{{Marcel} et~al.,}{{Marcel}
  et~al.}{2022}]{marcel22}
{Marcel} G.,  et~al., 2022, \mn@doi [\aap] {10.1051/0004-6361/202141375}, \href
  {https://ui.adsabs.harvard.edu/abs/2022A&A...659A.194M} {659, A194}

\bibitem[\protect\citeauthoryear{{Miller}, {Raymond}, {Fabian}, {Steeghs},
  {Homan}, {Reynolds}, {van der Klis}  \& {Wijnands}}{{Miller}
  et~al.}{2006}]{miller06}
{Miller} J.~M.,  {Raymond} J.,  {Fabian} A.,  {Steeghs} D.,  {Homan} J.,
  {Reynolds} C.,  {van der Klis} M.,   {Wijnands} R.,  2006, \mn@doi [\nat]
  {10.1038/nature04912}, \href
  {https://ui.adsabs.harvard.edu/abs/2006Natur.441..953M} {441, 953}

\bibitem[\protect\citeauthoryear{{Miller} et~al.,}{{Miller}
  et~al.}{2016}]{miller16}
{Miller} J.~M.,  et~al., 2016, \mn@doi [\apjl] {10.3847/2041-8205/821/1/L9},
  \href {https://ui.adsabs.harvard.edu/abs/2016ApJ...821L...9M} {821, L9}

\bibitem[\protect\citeauthoryear{{Motta} et~al.,}{{Motta}
  et~al.}{2021}]{motta21}
{Motta} S.~E.,  et~al., 2021, \mn@doi [\nar] {10.1016/j.newar.2021.101618},
  \href {https://ui.adsabs.harvard.edu/abs/2021NewAR..9301618M} {93, 101618}

\bibitem[\protect\citeauthoryear{{Negoro} et~al.,}{{Negoro}
  et~al.}{2018}]{negoro18}
{Negoro} H.,  et~al., 2018, The Astronomer's Telegram, \href
  {https://ui.adsabs.harvard.edu/abs/2018ATel12254....1N} {12254, 1}

\bibitem[\protect\citeauthoryear{{Oeda} et~al.,}{{Oeda} et~al.}{2019}]{oeda19}
{Oeda} .,  et~al., 2019, The Astronomer's Telegram, \href
  {https://ui.adsabs.harvard.edu/abs/2019ATel12398....1O} {12398, 1}

\bibitem[\protect\citeauthoryear{{Petrucci}, {Ferreira}, {Henri}, {Malzac}  \&
  {Foellmi}}{{Petrucci} et~al.}{2010}]{pop10}
{Petrucci} P.~O.,  {Ferreira} J.,  {Henri} G.,  {Malzac} J.,   {Foellmi} C.,
  2010, \mn@doi [\aap] {10.1051/0004-6361/201014753}, \href
  {https://ui.adsabs.harvard.edu/abs/2010A&A...522A..38P} {522, A38}

\bibitem[\protect\citeauthoryear{{Petrucci} et~al.,}{{Petrucci}
  et~al.}{2021}]{pop21}
{Petrucci} P.~O.,  et~al., 2021, \mn@doi [\aap] {10.1051/0004-6361/202039524},
  \href {https://ui.adsabs.harvard.edu/abs/2021A&A...649A.128P} {649, A128}

\bibitem[\protect\citeauthoryear{{Pinto}, {Middleton}  \& {Fabian}}{{Pinto}
  et~al.}{2016}]{pinto16}
{Pinto} C.,  {Middleton} M.~J.,   {Fabian} A.~C.,  2016, \mn@doi [\nat]
  {10.1038/nature17417}, \href
  {https://ui.adsabs.harvard.edu/abs/2016Natur.533...64P} {533, 64}

\bibitem[\protect\citeauthoryear{{Pinto} et~al.,}{{Pinto}
  et~al.}{2021}]{pinto21}
{Pinto} C.,  et~al., 2021, \mn@doi [\mnras] {10.1093/mnras/stab1648}, \href
  {https://ui.adsabs.harvard.edu/abs/2021MNRAS.505.5058P} {505, 5058}

\bibitem[\protect\citeauthoryear{{Ponti}, {Fender}, {Begelman}, {Dunn},
  {Neilsen}  \& {Coriat}}{{Ponti} et~al.}{2012}]{ponti12}
{Ponti} G.,  {Fender} R.~P.,  {Begelman} M.~C.,  {Dunn} R.~J.~H.,  {Neilsen}
  J.,   {Coriat} M.,  2012, \mn@doi [\mnras]
  {10.1111/j.1745-3933.2012.01224.x}, \href
  {https://ui.adsabs.harvard.edu/abs/2012MNRAS.422L..11P} {422, L11}

\bibitem[\protect\citeauthoryear{{Russell} et~al.,}{{Russell}
  et~al.}{2022}]{russell22}
{Russell} T.~D.,  et~al., 2022, \mn@doi [\mnras] {10.1093/mnras/stac1332},
  \href {https://ui.adsabs.harvard.edu/abs/2022MNRAS.513.6196R} {513, 6196}

\bibitem[\protect\citeauthoryear{{Tetarenko}, {Sivakoff}, {Heinke}  \&
  {Gladstone}}{{Tetarenko} et~al.}{2016}]{tetarenko16}
{Tetarenko} B.~E.,  {Sivakoff} G.~R.,  {Heinke} C.~O.,   {Gladstone} J.~C.,
  2016, \mn@doi [\apjs] {10.3847/0067-0049/222/2/15}, \href
  {https://ui.adsabs.harvard.edu/abs/2016ApJS..222...15T} {222, 15}

\bibitem[\protect\citeauthoryear{{Tomaru}, {Done}, {Ohsuga}, {Nomura}  \&
  {Takahashi}}{{Tomaru} et~al.}{2019}]{tomaru19}
{Tomaru} R.,  {Done} C.,  {Ohsuga} K.,  {Nomura} M.,   {Takahashi} T.,  2019,
  \mn@doi [\mnras] {10.1093/mnras/stz2738}, \href
  {https://ui.adsabs.harvard.edu/abs/2019MNRAS.490.3098T} {490, 3098}

\bibitem[\protect\citeauthoryear{{Tomaru}, {Done}  \& {Mao}}{{Tomaru}
  et~al.}{2023}]{Tomaru2023}
{Tomaru} R.,  {Done} C.,   {Mao} J.,  2023, \mn@doi [\mnras]
  {10.1093/mnras/stac3210}, \href
  {https://ui.adsabs.harvard.edu/abs/2023MNRAS.518.1789T} {518, 1789}

\makeatother
\end{thebibliography}









\newpage

\label{lastpage}
\end{document}